\RequirePackage{fix-cm}
\documentclass[twocolumn,epjc3,showpacs,preprintnumbers,amsmath,amssymb]{svjour3}
\smartqed
\RequirePackage{graphicx}
\RequirePackage[numbers,sort&compress]{natbib}
\usepackage[colorlinks,linkcolor=blue,anchorcolor=blue,citecolor=green]{hyperref}
\usepackage{graphicx,amsmath,amsfonts,latexsym,amssymb,graphics,epsfig,subfigure,color,makeidx}
\usepackage{multirow,epstopdf}
\journalname{Eur. Phys. J. C}

\begin{document}

\title{Fundamental energy scale of the thick brane in mimetic gravity}

\author{Tao-Tao Sui\thanksref{addr1,addr2,addr3}
                \and Yu-Peng Zhang\thanksref{addr1,addr2,addr3}
             \and Bao-Min Gu\thanksref{addr4}
             \and Yu-Xiao Liu\thanksref{e4,addr1,addr2,addr3}
        }

\thankstext{e4}{e-mail:liuyx@lzu.edu.cn, corresponding author}
   \institute{Institute of Theoretical Physics $\&$ Research Center of Gravitation, Lanzhou University, Lanzhou 730000, China  \label{addr1}
  \and
   Joint Research Center for Physics, Lanzhou University and Qinghai Normal University, Lanzhou 730000 and Xining ,810000, China \label{addr2}
   \and
   Lanzhou Center for Theoretical Physics, Lanzhou University, Lanzhou, Gansu 730000, China \label{addr3}
   \and
  Department of Physics, Nanchang University, Nanchang 330031, China \label{addr4}
}

\date{Received: date / Accepted: date}

 \maketitle

\begin{abstract}
In this paper, thick branes generated by the mimetic scalar field with Lagrange multiplier formulation are investigated. We give three typical thick brane background solutions with different asymptotic behaviors and show that all the solutions are stable under tensor perturbations. The effective potentials of the tensor perturbations exhibit as volcano potential, Po\"{o}schl-Teller potential, and harmonic oscillator potential for the three background solutions, respectively. All the tensor zero modes (massless gravitons) of the three cases can be localized on the brane. We also calculate the corrections to the Newtonian potential. On a large scale, the corrections to the Newtonian potential can be ignored. While on a small scale, the correction from the volcano-like potential is more pronounced than the other two cases. Combining the specific corrections to the Newtonian potential of these three cases and the latest results of short-range gravity experiments, we get the constraint on the scale parameter as $k\gtrsim 10^{-4}$eV, and constraint on the corresponding five-dimensional fundamental scale as $M_\ast \gtrsim10^5$TeV.
\end{abstract}

\maketitle

\section{Introduction}

{The nature of dark matter is one of the most mysterious fields in the modern theoretical physics \cite{planck,boss}. Over the last century, the theoretical physicists have made many attempts to reveal the mask of dark matter \cite{boss1,boss2,boss3,boss4,boss5,boss6}.  One possible approach is that the general relativity should be modified at large scale \cite{modegravity1,modegravity2,modegravity3,modegravity4,modegravity5}. Recently, a Weyl-symmetric extension of general relativity has attracted more and more attention, also named as mimetic gravity \cite{Chamseddine:2013kea,Chamseddine:2014vna}.}

In the mimetic gravity model, {the physical metric $g_{\mu\nu}$ is determined by an auxiliary metric $\tilde{g}_{\mu\nu}$ and a scalar field $\phi$ with the relation of $g_{\mu\nu}=-\tilde{g}_{\mu\nu}\tilde{g}^{\alpha\beta}\partial_\alpha\phi\partial_\beta\phi$ \cite{Chamseddine:2013kea}.} In such a setup, the conformal degree of freedom is separated with a covariant way, and the extra degree of freedom can be deemed to dynamic and mimic cold dark matter. The mimetic model could be transformed into a Lagrange multiplier formulation with a potential of the mimetic scalar field. With these methods, one can obtain a viable theory confronted with the cosmic evolution. It was pointed out that this model can also drive the late-time acceleration and early-time inflation \cite{Momeni:2015gka}. {In Ref. \cite{aoba}, the authors pointed out that the mimetic gravity is free of ghost instability only for a positive energy density. Then, Chaichian et al. proved that the mimetic gravity based on tensor-vector theory or tensor-vector-scalar theory is free of ghost instabilities \cite{chichian}.}  For more recent works about mimetic gravity, see Refs.~\cite{Matsumoto:2015wja,Oikonomou:2015lgy,Cognola:2016gjy,Rabochaya_2016,Odintsov:2016imq,Sebastiani:2016ras,Golovnev,Deruelle,Momeni,Leon,Myrzakulov,Mukhanov,Mukhanov2,Mukhanov1,Astashenok2015}.

\begin{figure}[htb]
\begin{center}
 {\label{thin}\includegraphics[width=0.3\textwidth]{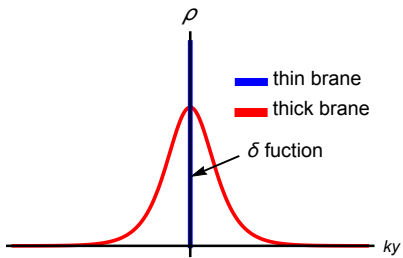}}
 \end{center}
\caption{The shapes of energy distributions of thin and thick branes.} \label{thinandthick}
\end{figure}

{On the other hand,  since the brane world theory can address the hierarchy problem and the cosmological constant problem successfully \cite{Randall:1999ee,Randall:1999vf,Kim:2000mc}, it has attracted more and more attention in the last decades.} In the brane world theory, our observable universe is supposed as a 3-brane embedded into a higher-dimensional bulk. The elementary particles in the standard model should be localized on the brane and gravity can propagate into the extra dimension. According to the energy distribution of the brane along the extra dimension, brane models can be divided into thin brane models and thick brane models. For a thin brane model, the energy distribution is a Dirac delta function \cite{Davoudiasl:1999tf,Shiromizu:1999wj,Gherghetta:2000qt,Rizzo:2010zf,Yang:2012dd,Agashe:2014jca,keyang}. For a thick brane model, the energy distributes along the extra dimension but localizes around some narrow regions \cite{Csaki:2000fc,DeWolfe:1999cp,Gremm:1999pj,Liu:2017gcn,Afonso:2006gi,Afonso:2007zz,Guo:2014bxa,German:2013sk,Liu:2009ega,Dzhunushaliev:2009va,Arias:2002ew,BarbosaCendejas:2006hj,Liu:2011am,Dzhunushaliev:2009dt,Zhong:2015pta,Bronnikov:2020}. Figure \ref{thinandthick} shows the shapes of thin brane \cite{Randall:1999vf} and thick brane \cite{gubaomin}.

It is known that there is no dynamic for a thin brane. For investigating the dynamical generation of a brane and its internal structure, thick brane models were presented. A typical mechanism is that the thick brane can be generated by one or more background scalar fields coupled with gravity. It means that the features of the background scalar fields can influence the construction of the thick brane, namely, different scalar field can generate thick brane with  different structure.

Recently, in Refs. \cite{zhonyi1,guowendi}, the authors promoted the four-dimensional mimetic gravity into the brane world theory. The five-dimensional mimetic field $\phi$ can be regarded as a background scalar field which can generate a thick brane. According to this setup, they investigated some thick brane models with more multi sub-branes. Along the direction of the extra dimension, the Kaluza-Klein (KK) graviton modes are trapped in a volcano-like effective potential.

{Note that, once we consider the brane world theory, a series of massive particles beyond the standard model of particle physics will be predicted, e.g., massive gravitons and massive vector particles. These massive particles will correct the forms of the four-dimensional Newtonian potential and Coulomb potential. For example, a massive graviton will contribute a correction term to the four-dimensional Newtonian potential with a form of Yukawa potential, and the corresponding total contribution depends on the mass spectrum of the gravitons.

Furthermore, many researches on the brane world theory result that the effective potentials, which are the KK graviton modes trapped in, depend on the structures of brane world. It means that the different brane world structures will lead to different mass spectra of the KK gravitons, and these differences will eventually be reflected in the correction behavior to the Newtonian potential.

According to Refs. \cite{zhonyi1,guowendi}, we can see that the effective potential of the tensor perturbation for the mimetic brane model with Lagrange multiplier formulation only depend on the warp factor $A(z)$
\begin{equation}
V_{eff}=\frac{3}{2}\partial_{z}^{2}A+\frac{9}{4}(\partial_z A)^2.
\end{equation}
In other words, different warp factors can result different correction behaviors to the Newtonian potential.  Besides, the localization of the massless graviton mode requires that for $z\rightarrow\infty$, the effective potential $V_{eff}$ should satisfy $V_{eff}|_{z\rightarrow\infty}\ge0$. There are three asymptotic behaviors of the effective potentials satisfy the requirement,
\begin{equation}\label{asympotic}
V_{eff}|_{z\rightarrow\infty}\Rightarrow\begin{cases}
0,\\
\text{constant},\\
\infty.
\end{cases}
\end{equation}
The corresponding effective potentials are volcano-like,  P\"{o}schl-Teller-like, and harmonic-oscillator-like effective potentials. Note that the four-dimensional mimetic gravity model can result in an extra degree of freedom to explain dark matter, the existence of such degree of freedom relax the constraint for constructing a thick brane, therefore we can construct three different brane world models which correspond above three effective potential by considering the higher-dimensional mimetic model. We can compare the difference of the correction behaviors to the four-dimensional Newtonian potential caused by the three different brane world structures. Then, by combining the current gravitational inverse-square law experiments \cite{luojun,luojun1,Lee}, we can obtain the fundamental energy of the brane world in mimetic gravity.}

The organization of this paper is as follows. In Sec.~\ref{Sec2}, we briefly introduce the mimetic theory in the brane world scenario and obtain the corresponding thick brane solutions. Then, we analyze the stability of the brane solutions under the tensor perturbations and check the localization of the massless graviton in Sec.~\ref{Sec3}.  After that, we derive the corresponding correction to the four-dimensional Newtonian potential, for which the final constraints about the mimetic thick brane are given in Sec.~\ref{Sec4}. Finally, the conclusion and discussion are given in Sec.~\ref{Sec5}.

\section{THE MODEL} \label{Sec2}

In this section, we consider the five-dimensional mimetic gravity with the following action
 \begin{eqnarray}
        S=\int d^5x\sqrt{-g}\left( \frac{R}{2\kappa^2_5}
        + \lambda\left[\partial_M \phi \partial^M \phi-U\right]-V(\phi) \right),
        \label{action mgb1}
    \end{eqnarray}
where $\kappa^2_5=1/M_\ast^3$ with $M_\ast$ being the five-dimensional fundamental scale and $\lambda$ is a Lagrange multiplier. For simplicity, we chose the natural unit with $\kappa^2_5=1$.

{In this paper, we set the thick brane to be static and flat. In the brane world theory, the thick brane can be generated by a five-dimensional background scalar field. On the one hand, for a static thick brane, the background filed $\phi$ should not be a function of time. On the other hand, for a flat brane, the four-dimensional Lorentz symmetry should be satisfied on the brane. Therefore, the scalar field $\phi$ should only depend on the extra dimension, which means that there is no kinetic energy term of $\phi$ and the {mimetic} thick brane is free of ghost instability. In such setup, we can see that $U=g^{MN}\partial_{M}\phi\partial_{N}\phi=e^{-2A}\big(\partial_{z}\phi\big)^2>0$}.

By varying the action \eqref{action mgb1} with respect to $g_{MN}$, $\phi$, and $\lambda$, respectively, we get the equations of motion as
    \begin{eqnarray}
        \label{var eom1}
        G_{MN}+2\lambda \partial_M \phi \partial_N \phi-L_{\phi}g_{MN}=0, \\
        \label{var eom2}
        2\lambda\Box^{(5)}\phi+2g^{MN}\partial_{M}\lambda\partial_{N}\phi+\lambda \frac{\partial U}{\partial \phi}+\frac{\partial V}{\partial \phi}=0, \\
        \label{var eom3}
        g^{MN}\partial_M \phi \partial_N \phi-U=0.
    \end{eqnarray}
Here, $L_{\phi}=\lambda\left[g^{MN}\partial_M \phi \partial_N \phi-U\right]-V(\phi)$, and the five-dimensional d'Alembert operator is defined as  $\Box^{(5)}=g^{MN}\nabla_{M}\nabla_{N}$. The Latin indices $(M, N = 0, 1, 2, 3, 5)$ stand for
the five-dimensional coordinate indices, and the Greek indices $(\mu, \nu = 0, 1, 2, 3)$ represent the brane coordinate indices.

We consider the following {brane-world} metric with four-dimensional Poincar\'{e} invariance
    \begin{eqnarray}
        \label{brane metric1}
       ds^2=e^{2A(z)}\Big(\eta_{\mu\nu}dx^{\mu}dx^{\nu}+dz^2\Big),
    \end{eqnarray}
{where $e^{A(z)}$ or ${A(z)}$ is called as the warp factor.}
Then, Eqs. (\ref{var eom1})-(\ref{var eom3}) can be rewritten as
    \begin{eqnarray}
        \label{eom21}
      e^{-2A}(3A'^2+3A'')+V(\phi)+\lambda\Big(U-e^{-2A}\phi'^2\Big)=0, \\
       \label{eom22}
       6A'^2+e^{2A}V(\phi)+e^{2A}\lambda \Big(U+e^{-2A}\phi'^2\Big)=0, \\
       \label{eom23}
       \lambda \left(6A'\phi'+2\phi''+e^{2A}\frac{\partial U}{\partial\phi}\right)
        +2\lambda'\phi'+e^{2A}\frac{\partial V}{\partial \phi}=0,  \\
       \label{eom24}
       e^{-2A}\phi'^2-U=0,
    \end{eqnarray}
where the prime denotes the derivative with respect to the extra-dimensional coordinate $z$. The above equations are not independent of each other. Combining them we get three largely simplified equations for $\lambda$, $U(\phi)$, and $V(\phi)$
\begin{eqnarray}
\label{sol lambda}
\lambda &=&\frac{3(A''-A'^2)}{2\phi'^2},\\
\label{sol U}
U&=&e^{-2A} {\phi'^2},\\
\label{sol V}
V(\phi) &=&-3e^{-2A}(A'^2+A'').
\end{eqnarray}

{Generally, in order to get the solutions of thick brane, we can use the {super-potential} approach by setting a related variable for the brane model with two independent equations and three variables, more details can refer Refs. \cite{DeWolfe:1999cp,Bazeia1q}. In this model, we get three independent equations, and five variables e.g, $A(z),\phi,\lambda,V(\phi)$, and $U$. Inspired by the {super-potential} approach, we need to constraint the two variables of them. Note that all the expressions of $\lambda$, $U$, and $V(\phi)$ depend on the warp factor $A(z)$ and the mimetic field $\phi(z)$. So, once $A(z)$ and $\phi(z)$ are given, the profiles of $\lambda$, $U$, and $V(\phi)$ could be determined. Besides, we can see that only the variable $A(z)$ can affect the perturbed tensor equation (31).} On the one hand, there are no constraints on $A(z)$ and $\phi(z)$ from the equations of motion, so they can be chosen arbitrarily in principle. However, on the other hand, a viable thick {brane-world} model should satisfy the minimal requirement, namely, the localization of massless tensor mode (massless graviton). Therefore, not all the choices of $A(z)$ and $\phi(z)$ are achievable. Here, we will consider three typical solutions.

\subsection{Volcano (VO) type thick brane}

Firstly, we consider the case of {the warp factor} $e^{A(z)}$ as a power function of the extra-dimensional $z$, and the mimetic field $\phi(z)$ is a kink function. The solutions can be given as
\begin{eqnarray}
\label{case1A}
e^{A(z)}=\frac{1}{\sqrt{k^2z^2+1}},~ \phi(z)=v\left(\frac{kz}{\sqrt{1+k^2z^2}}\right)^\gamma,
\end{eqnarray}
where $k$ is the scale parameter which controls the thickness of the brane,  $\gamma$ is a positive integer, and $v$ is a positive parameter {determining} the limit of the scalar field. The corresponding expressions of  $\lambda$, $U(\phi)$, and $V(\phi)$ can be expressed as
\begin{eqnarray}
\label{sol lambda1}
\lambda &=&-\frac{3}{2}\frac{(k^2z^2+1)^\gamma}{\gamma^2v^2(k^2z^2)^{\gamma-1}},\\
\label{sol U1}
U&=&\frac{\gamma^2v^2}{z^2(1+k^2z^2)}\left(\frac{kz}{\sqrt{1+k^2z^2}}\right)^{2\gamma},\\
\label{sol V1}
V(\phi) &=&-3k^2(3\Phi^{{2}/{\gamma}}-1),
\end{eqnarray}
where $\Phi=\phi/v$. {Such brane solution will give a volcano type effective potential of the tensor perturbations.}

\subsection{P\"{o}schl-Teller (PT) type thick brane}

{Then, we come to the hyperbolic function form of the warp factor $e^{A(z)}$ and a different kink form of the mimetic field $\phi(z)$.} The expressions of warp factor and mimetic field can be given as
\begin{eqnarray}
\label{case2A}
e^{A(z)}=\text{sech}(kz), ~\phi(z)=v\text{tanh}^\gamma(kz),
\end{eqnarray}
{for which} the other {functions} can be {solved} as
\begin{eqnarray}
\label{sol lambda2}
\lambda &=&-\frac{3 \sinh^2(2 k z) \tanh^{-2 n}(k z)}{8 \gamma^2 v^2},\\
\label{sol U2}
U&=&\Big(k \gamma v~\text{csch}(k z)\tanh(k z)\Big)^2,\\
\label{sol V2}
V(\phi) &=&\frac{3 k^2 \left(2 \Phi ^{{2}/{\gamma}}-1\right)}{\Phi ^{{2}/{\gamma}}-1}.
\end{eqnarray}

\subsection{Harmonic oscillator (HO) type thick brane}

{Finally, we choose an exponential warp factor and a kink mimetic field:}
\begin{eqnarray}\label{case3A}
e^{A(z)}=e^{-k^2z^2}, ~\phi(z)=v(\frac{kz}{\sqrt{1+k^2z^2}})^\gamma.
\end{eqnarray}
The specific expressions of $\lambda$, $U(\phi)$, and $V(\phi)$ {are solved as follows}
\begin{eqnarray}
\label{sol lambda3}
\lambda &=&-\frac{3 \left(2 k^2 z^2+1\right) \left(k^2 z^2+1\right)^{\gamma +2}}{\gamma ^2 v^2 \left(k^2 z^2\right)^{\gamma -2}},\\
\label{sol U3}
U&=&\Big(\frac{e^{k^2z^2}\gamma v}{z(1+k^2z^2)}\Big)^2(\frac{kz}{\sqrt{1+k^2z^2}})^{2\gamma},\\
\label{sol V3}
V(\phi) &=&\frac{6 k^2 \left(3 \Phi ^{v }-1\right) }{\Phi ^{{2}/{\gamma}}-1}
           e^{\frac{2}{\Phi ^{-{2}/{\gamma}}-1}}.
\end{eqnarray}

\begin{figure*}[!htb]
\begin{center}
  \subfigure[~$\phi$ for the cases VO and HO]{\label{casephi1}\includegraphics[width=0.30\textwidth]{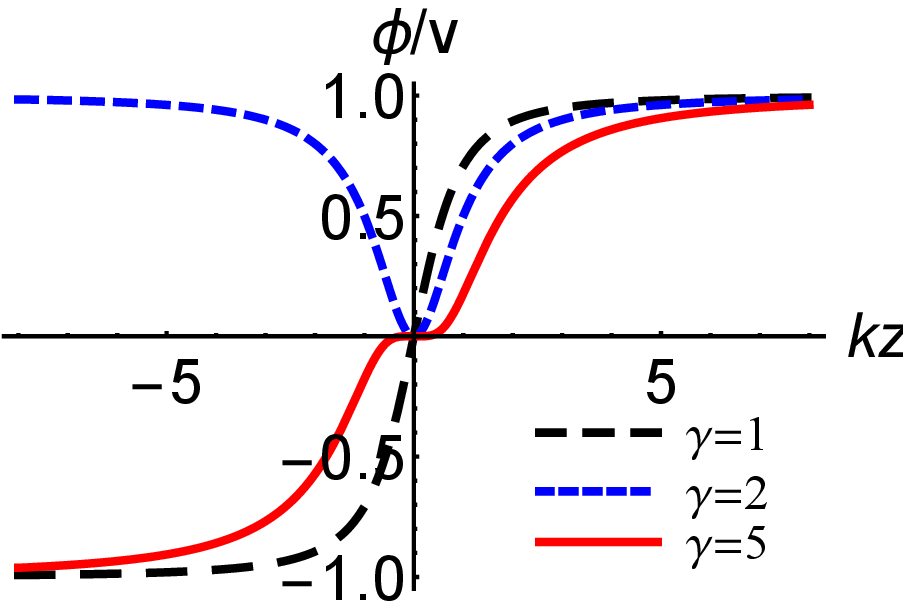}}
  \subfigure[~$\phi$ for the case PT]{\label{casephi2}\includegraphics[width=0.30\textwidth]{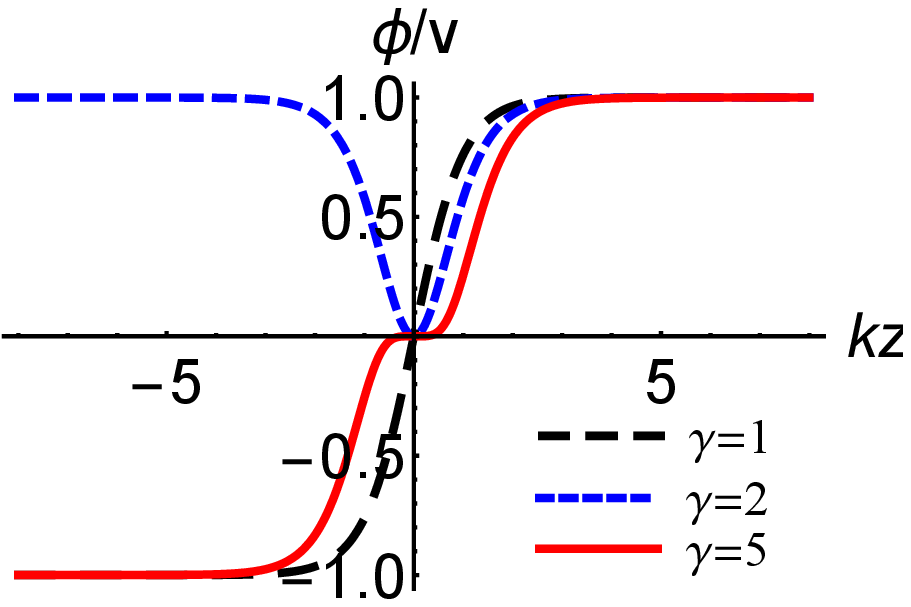}}
  \subfigure[~$e^A$ for all cases]{\label{casewarp}\includegraphics[width=0.30\textwidth]{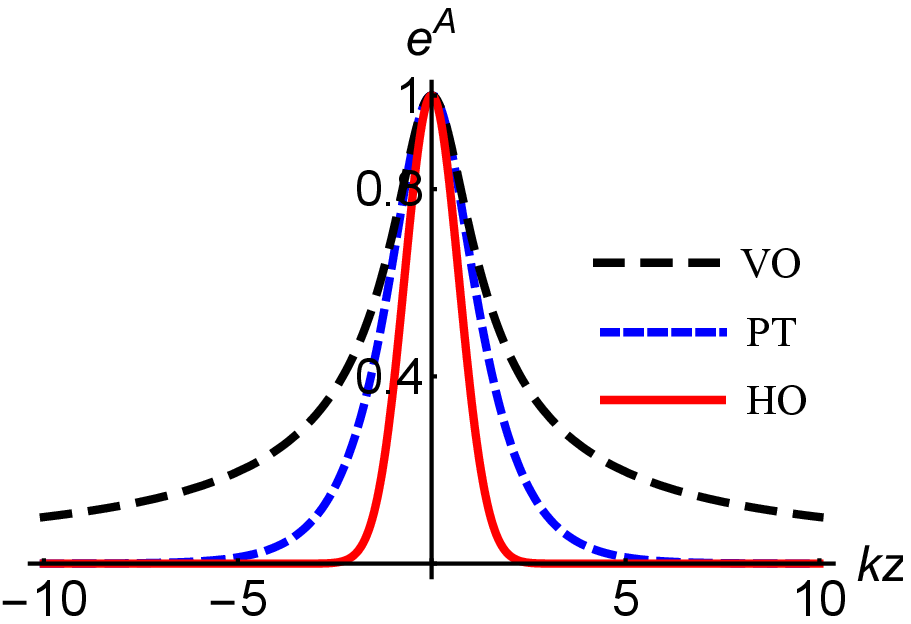}}
 \end{center}
\caption{The shapes of  the mimetic field $\phi(z)$ and the warp factors $e^{A}$ for all cases.} \label{solution1}
\end{figure*}

The shapes of these three kinds of warp factors and the two kinds of mimetic scalar fields are shown in Fig. \ref{solution1}.  Figures \ref{casephi1} and \ref{casephi2} show that if $\gamma$ is an odd integer, the mimetic scalar field would be a single-kink (the black dashed lines) for $\gamma=1$, and it will become a double-kink (the red lines) with $\gamma\geq3$. Besides, if $\gamma$ is an even integer, the scalar field will be  not a kink configuration (the blue dashed line) anymore. For a general thick brane model, the background scalar field should be a kink configuration. While, in mimetic thick brane model, due to the Lagrange multiplier which can cause excess degrees of freedom, the non-kink scalar field can also generate a thick brane. Although these solutions have the same limit of $e^{A}|_{z\rightarrow\infty}\rightarrow 0$, they differ with {the asymptotic behaviors at infinity of the extra dimension}, with the attenuation intensity $\text{HO}>\text{PT}>\text{VO}$ (see Fig. \ref{casewarp}). These different asymptotic behaviors will lead to different physical properties, including the potentials felt by the gravitons along the extra dimension and the corrections of the Newtonian potential caused by the massive gravitons.

\section{linear perturbations and localization} \label{Sec3}

{In this section, we consider the linear perturbations of the metric and their localization. It is well known that the linear perturbations of a background metric can be decomposed into three parts: the transverse-traceless tensor modes, the scalar modes, and the transverse vector modes (the so-called scalar-vector-tensor (SVT) decomposition), for which the three kinds of modes decouple with each other \cite{massimo}. }

{According to the Bardeen formalism for metric fluctuations \cite{Bardeen}, the extra dimension part of the metric fluctuations can be expressed as a scalar mode. Besides, due to the SVT decomposition, the tensor and scalar modes are decoupled. In the brane world theory, the tensor modes of perturbations can describe the gravitons which can influence the four-dimensional Newtonian potential. Then, combining the short-range experiments, we can obtain the constraint of the mimetic gravity. Therefore, we take the form of the tensor perturbations as follows}
\begin{eqnarray}
        \label{tensor metric}
        ds^2=e^{2A(z)}\Big[(\eta_{\mu\nu}+h_{\mu\nu})dx^{\mu}dx^{\nu}+dz^2\Big].
\end{eqnarray}
Here, $h_{\mu\nu}=h_{\mu\nu}(x^{\mu},z)$ depends on all the coordinates.
Combining the specific perturbed metric \eqref{tensor metric} and the transverse-traceless (TT) condition, i.e., $\partial^\mu h_{\mu\nu}=\eta^{\mu\nu}h_{\mu\nu}=0$, we can simplify the perturbed Ricci tensor as
\begin{eqnarray}
\delta R_{\mu\nu} \!\!&=& \!\! -\frac{1}{2}\Big(\Box^{(4)}+\partial^2_z+2A''+6A'^2+3A'\partial_z\Big)h_{\mu\nu},
~~~~~ \label{perturbed Ricci2}\\
\delta R_{55} \!\! &=& \!\! 0,~~~\label{perturbed Ricc2}
\end{eqnarray}
where the four-dimensional d'Alembertian is defined as $\Box^{(4)}\equiv\eta^{\mu\nu}\partial_{\mu}\partial_{\nu}$.
Besides, the tensor perturbations of the equations of motion \eqref{var eom1} can be expressed as the following form
\begin{eqnarray}\label{eome2}
\delta R_{MN}=\frac{2}{3}\delta g_{MN}(\lambda U+V).
\end{eqnarray}
Then, by combining Eqs.~\eqref{sol lambda}, \eqref{sol U}, \eqref{sol V}, and the perturbed metric \eqref{tensor metric}, {the right hand side of Eq.~\eqref{eome2} can be simplified as}
\begin{eqnarray}
\frac{2}{3}\delta g_{\mu\nu}(\lambda U+V) &=& -(A''+3A'^2)h_{\mu\nu},\\
\frac{2}{3}\delta g_{55}(\lambda U+V)&=& 0.
\end{eqnarray}
Therefore, the perturbed tensor equation can be obtained as
\begin{equation}\label{perturbed tensor equation}
-\frac{1}{2}\Box^{(4)}h_{\mu\nu}-\frac{1}{2}h_{\mu\nu}''
-\frac{3}{2}A'h_{\mu\nu}'=0.
\end{equation}

Next, we make a KK decomposition $\tilde{h}_{\mu\nu}=\sum\epsilon^{(n)}_{\mu\nu}(x)\tilde{\Psi}_{n}(z)$, where the polarization tensor $\epsilon_{\mu\nu}$ also satisfies the TT condition $\partial^\mu \epsilon_{\mu\nu}=\epsilon_{\mu}^{~\mu}=0$. Bringing the KK decomposition into the perturbed tensor equation \eqref{perturbed tensor equation}, we can get a four-dimensional massive Klein-Gordon equation for the polarization tensor $\epsilon_{\mu\nu}(x)$ and an equation for the extra-dimensional part $\tilde{\Psi}_{n}(z)$:
\begin{eqnarray}
\Big(\Box^{(4)}-m^{2}_{n}\Big)\epsilon^{(n)}_{\mu\nu}(x)&=&0,\label{KGepsilon}\\
-\tilde{\Psi}_{n}''(z)-3A'\tilde{\Psi}_{n}'(z)&=&m^2_{n}\tilde{\Psi}_{n}(z)\label{KGlambda}.
\end{eqnarray}
Furthermore, by redefining the extra dimensional part as $\tilde{\Psi}_{n}(z)=e^{-\frac{3}{2}A}\Psi_{n}(z)$, we obtain a Schr\"{o}dinger-like equation for the new function $\Psi_{n}(z)$ of the extra-dimensional part:
 \begin{eqnarray}\label{eq tensor}
-\partial^2_{z}\Psi_{n}(z)+V_{t}(z)\Psi_{n}(z)=m_n^2\Psi_{n}(z),
    \end{eqnarray}
with the effective potential $V_t(z)$ {given by}
\begin{eqnarray}\label{tensorpotential}
 V_t(z)=\frac{3}{2}A''+\frac{9}{4}A'^2.
\end{eqnarray}

We should note that with this effective potential \eqref{tensorpotential},  the above Schr\"{o}dinger-like equation can be rewritten as $\mathcal{H}\Psi_{n}(z)=m^2_{n}\Psi_{n}(z)$, where the Hamiltonian operator is given as $\mathcal{H}=Q^{+}Q$, with $Q=-\partial_{z}+\frac{3}{2}\partial_{z}A$.  Since the eigenvalue of the  Hamiltonian operator $\mathcal{H}$ is positive definite,  there do not exist negative $m_n^2$ modes, namely, there are no tachyonic tensor modes.

The abstract expression of the effective potential $V_{t}(z)$ shows that it only depends on the warp factor $A(z)$, namely, different asymptotic behaviors of the warp factor $A(z)$ can result in different properties of the effective potentials. We will discuss the properties of the effective potentials with the warp factors $A(z)$ given in Sec. \ref{Sec2}.

{Shapes of three kinds of effective potentials $V_t(z)$ are shown in Fig. \ref{tensoreffective}.}
From Fig. \ref{figure tensor11}, we can see that the effective potential $V_{t}(z)$ of case VO is a volcano-like potential with a single potential well, and the potential well becomes narrower and deeper with the parameter $k$ increases. The asymptotic behavior of the volcano-like potential shows that there are no other bound states except the zero mode, and the mass spectrum of the massive excited states is continuous from $m_n>0$.  Figure~\ref{figure tensor12} describes the shape of effective potential $V_{t}(z)$ for case PT. It shows that the second warp factor $A(z)$ leads to a PT potential behavior as $V_{t}(z)|_{z\rightarrow\pm\infty}={9k^2}/{4}$. The scale parameter $k$ determines the width and the depth of the potential well. For this PT effective potential, there are two bound states, i.e., the zero mode with $m_0=0$ and the first excited state with $m_1=\sqrt{2}k$, and the mass spectrum is also continuous from $m_n\geq\sqrt{2}k$. For case HO, Fig. \ref{figure tensor13} shows that the effective potential $V_{t}(z)$ has the behavior of a harmonic-oscillator potential with the parameter $k$ controls {the {the mass spectrum}. For the harmonic-oscillator potential, all of the states are bound states with mass $m_n=\sqrt{6n}k$, the index $n$ means $n-$th eigenstate. These different potentials lead to different mass spectra of massive gravitons, which can result in different corrections to the Newtonian potential.

At the end of this section, we consider the zero mode of the tensor perturbations by setting $m^2_{n}=0$ in Eq. \eqref{eq tensor}. It is easy to get the solution of the zero mode:
\begin{eqnarray}
\Psi_0 (z)\propto e^{\frac{3}{2}A(z)}.
\end{eqnarray}
One can verify that the zero modes for the above three different types of warp factors are square-integrable and hence all the zero modes are localized around the brane. Thus, the four-dimensional Newtonian potential can be realized on the brane.

\begin{figure*}[!htb]
\begin{center}
 \subfigure[~Case VO]{\label{figure tensor11}\includegraphics[width=0.30\textwidth]{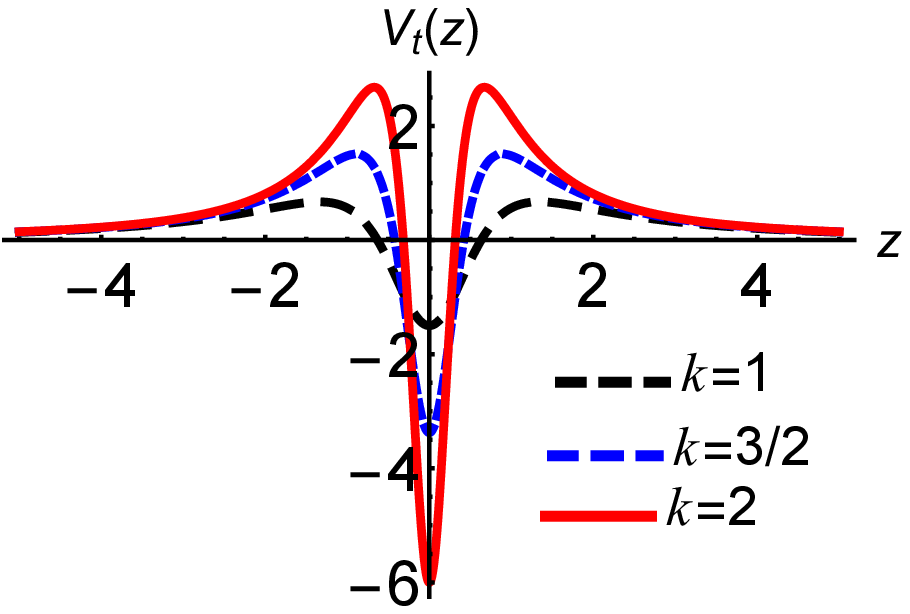}}
 \subfigure[~Case PT]{\label{figure tensor12}\includegraphics[width=0.30\textwidth]{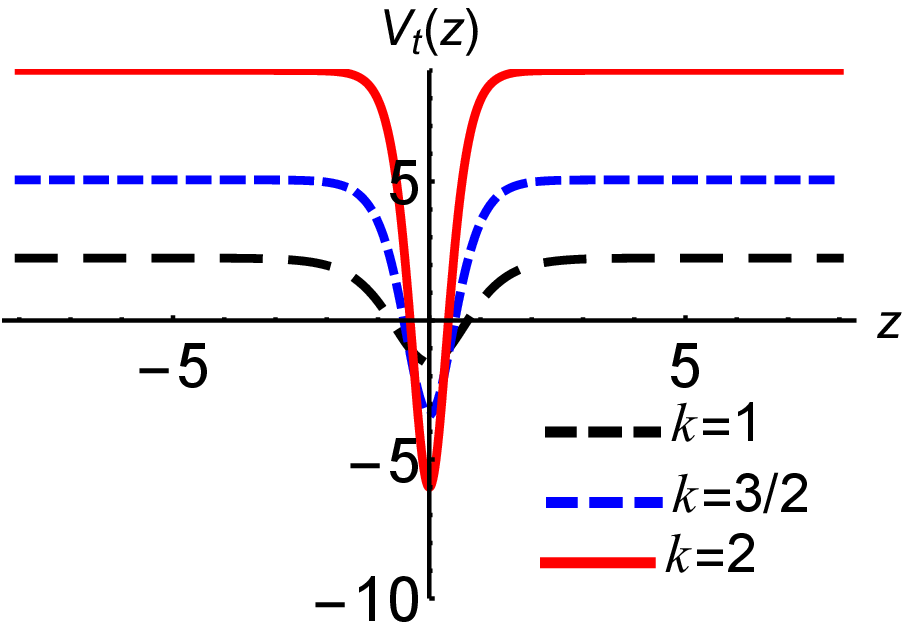}}
 \subfigure[~Case HO]{\label{figure tensor13}\includegraphics[width=0.30\textwidth]{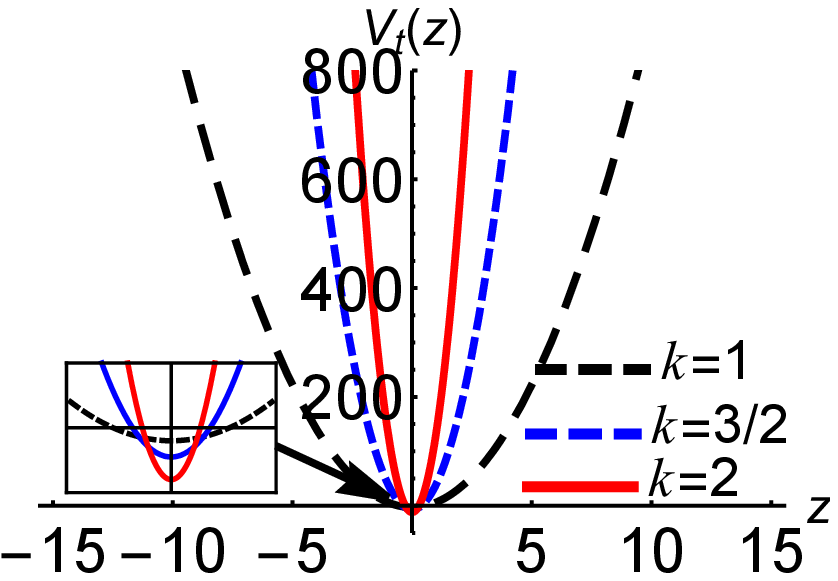}}
 \end{center}
\caption{The effective potentials  $V_t(z)$ of the tensor perturbations for all cases.} \label{tensoreffective}
\end{figure*}

\section{ The Correction to Newtonian Potential} \label{Sec4}
In the above section, we have considered the tensor perturbations and obtained the Schr\"{o}dinger-like equation \eqref{eq tensor}, we also demonstrated that the zero mode can be localized on the brane for all the cases to recover the four-dimensional gravity. {In this section}, we consider the massive KK modes of the gravitons which can cause the correction to the Newtonian potential in four-dimensional theory.

In the thick brane scenario, {the energy density of the brane has a distribution along the extra dimension. Therefore,} for simplicity, Refs. \cite{Rubakov, CsabaCsaki, arkanihamed3, Csaki:2000fc, Bazeia2009} considered the gravitational potential between two point-like sources of mass $M_1$ and $M_2$ located at the origin of the extra dimension, i.e., $z=0$. We can express the gravitational potential between two masses on the brane as
\begin{eqnarray}
V(r)&=-&\frac{M_1M_2}{M^{2}_{pl}}\frac{1}{r}-\frac{M_1M_2}{M^{3}_\ast}\sum \!\!\!\!\!\!\!\! {\int_{n\neq0}}\frac{e^{-mr}}{r}|\Psi_n(0)|^2\nonumber\\
&=&-\frac{M_1M_2}{M^{2}_{pl}}\frac{1}{r}\Big(1+\frac{M^{2}_{pl}}{M^{3}_\ast} \sum \!\!\!\!\!\!\!\! {\int_{n\neq0}}\Delta u(r)\Big),
\label{Newtonian_potential}
\end{eqnarray}
where $M_{pl}$ and $M_\ast$ are the effective four-dimensional Plank scale and the five-dimensional fundamental scale, respectively. $\sum \!\!\!\!\!\! {\int}\,_n$ stands for summation or integration (or both)  with respect to $n$, depending on the respective discrete or continuous character of the massive KK modes. Besides, we set $\Delta u(r)={e^{-mr}}|\Psi_n(0)|^2$.

We can focus on the curvature term of the action \eqref{action mgb1} from which we will derive the effective four-dimensional scale $M_{pl}$:
\begin{equation}\label{MplM1}
M^3_\ast \int d^5x\sqrt{-g}R\supset M_{pl}^2\int d^4x \sqrt{-g^{(4)}(x^\mu)}R^{(4)}(x^\mu).
\end{equation}
Therefore, the relation between the effective Planck scale $M_{pl}$ and the fundamental scale $M_\ast$ is given by
\begin{equation}\label{MplM2}
M_{pl}^2=M^3_\ast \int^{+\infty}_{-\infty}dz~e^{3A}.
\end{equation}
So, the gravitational potential between two masses on the brane can be simplified as
\begin{eqnarray}
V(r)&=&-\frac{M_1M_2}{M^2_{pl}}\Big(\frac{1}{r}+\Delta V(r)\Big)\nonumber\\
&=&-\frac{M_1M_2}{M^2_{pl}}\frac{1}{r}\Big(1+\Delta U(r)\Big),\label{Newtonian_potential0}\\
\Delta U(r)&=&\Big(\int^{+\infty}_{-\infty}dz~e^{3A}\Big) \sum \!\!\!\!\!\!\!\! {\int_{n\neq0}}\Delta u(r)\label{Newtonian_potential},
\end{eqnarray}
where $\Delta V(r)=\frac{1}{r}\Delta U(r)$ {is the correction term to the Newtonian potential, $\Delta U(r)$ the relative correction term and $\Delta u(r)$ the correction factor.} Next, we calculate the corrections to the Newtonian potential for the three cases.

\textbf{{Case VO:}}~We {substitute} the warp factor \eqref{case1A} into the Schr\"{o}dinger-like equation \eqref{eq tensor}, and get the reduced Schr\"{o}dinger-like equation
\begin{eqnarray}
-\partial^{2}_{z}\Psi_{n}+\frac{3 k^2 \left(5 k^2 z^2-2\right)}{4 \left(k^2 z^2+1\right)^2}\Psi_{n}=m^2_n\Psi_{n}.\label{case1reduce}
\end{eqnarray}
To solve {the above} equation, we consider the behavior of the effective potential {at infinity of the extra dimension:}
\begin{eqnarray}
V(z)\sim\frac{15}{4 z^2}.
\end{eqnarray}
The approximate solution is given by a linear combination of Bessel functions as
\begin{eqnarray}
\Psi_{n}(z)=\sqrt{z}\Big({C}_{1}\text{J}_{2}(m_nz)+{C}_{2}\text{Y}_{2}(m_nz)\Big),
\end{eqnarray}
where $C_1$ and $C_2$ are arbitrary constants, $\text{J}_{2}(m_nz)$ and $\text{Y}_{2}(m_nz)$ are the first and second Bessel functions, respectively.  In Ref. \cite{Csaki:2000fc}, the authors calculated the expression of $\Psi_{n}(0)$:
\begin{eqnarray}
\Psi_{n}(0)\sim (\frac{m_{n}}{k})^{1/2}.
\end{eqnarray}
So, the correction factor $\Delta u(r)$ to the Newtonian potential with a massive graviton $m_{n}$ is
\begin{eqnarray}
\Delta u(r)={e^{-m_nr}}|\Psi_n(0)|^2=\frac{m_n e^{-m_nr}}{k}.
\end{eqnarray}
From Fig. \ref{figure tensor11}, we can see that the spectrum of the massive gravitons is continuous. Therefore, the relative correction term to the Newtonian potential resulted by all the massive gravitons is
\begin{eqnarray}
\Delta U(r)&=& \Big(\int^{+\infty}_{-\infty}dz~e^{3A}\Big) \int_{0}^{\infty}dm\Delta u(r)\nonumber\\
&=&\frac{2}{k^2r^2}.
\end{eqnarray}
The correction to the Newtonian potential for case VO is the same form as that of the Randall-Sundrum brane model $\Delta V(r)= \frac{1}{r}\Delta U(r)\sim 1/k^2r^3$.

\textbf{{Case PT:}}~For this case with the warp factor \eqref{case2A}, the corresponding effective potential turns into
\begin{eqnarray}
V_t(z)=\frac{9}{4}k^2-\frac{15}{4}k^2\text{sech}^2 (kz),
\end{eqnarray}
and the Schr\"{o}dinger-like equation can be expressed as
\begin{eqnarray}
\Big(-\partial^2_z-\frac{15}{4}k^2\text{sech}^2(kz)\Big)\Psi_{n}=E_{n}\Psi_{n},
\end{eqnarray}
where $E_{n}=m^2_n-\frac{9}{4}k^2$. {It can be shown that there are two bound states in this potential. The first one is the ground state $\Psi_{0}(z)$ with $E_{0}=-\frac{9}{4}k^2$, and it is in fact the zero mode since the mass is zero: $m_0=0$. The second one is the first excited state $\Psi_{1}(z)$ with  $E_{1}=-k^2/4$, which represents a massive graviton with mass $m_{1}=\sqrt{2} k$.} The two bound states wave functions are
\begin{eqnarray}
\Psi_{0}(z)&=&C_{0}\text{sech}^{3/2}(kz),\\
\Psi_{1}(z)&=&C_{1}\text{sinh}(kz)\text{sech}^{3/2}(kz).
\end{eqnarray}
Here $C_{0}$ and $C_{1}$ are the normalization constants. We should note that $\Psi_{1}(0)=0$ which means that the  first excited state does not contribute to the correction of the Newtonian potential.

The continuous spectrum starts at $E_{n}=0$, corresponding to $m^{2}_{n}\geq 9k^2/4$. These excited states asymptotically turn into plane waves, and represent delocalized KK massive gravitons. Their explicit expressions can be given in terms of the associated Legendre functions of the first kind:
\begin{eqnarray}
\Psi_{n}(z)=\sum_{\pm} C^n_{\pm}P_{3/2}^{\pm \sigma}\text{tanh}(kz),
\end{eqnarray}
where $C^n_{\pm}$ are $m_{n}$-dependent parameters and
\begin{eqnarray}
\sigma=\sqrt{\frac{9}{4}k^2-m^2_n}.
\end{eqnarray}
In order to calculate the correction from the continuous modes, we need to compute the
normalization constants $C^n_{\pm}$. {According to} Refs. \cite{Cendejas2008,Fuentevilla2014,
Alfredo2013}, we can reduce the constants $C^n_{\pm}$ as
\begin{equation}
C^n_{+}=C^{n}_{-}=\frac{1}{\sqrt{2\pi}}|\Gamma(1+\sigma)|,
\end{equation}
{where $\Gamma(1+\sigma)$ is the gamma function.} So, $\Psi_{n}(0)$ can be expressed as the following form
\begin{equation}
\Psi_{n}(0)=\frac{\Gamma(1-\sigma)}{\Gamma(-\frac{1}{4}-\frac{\sigma}{2})
\Gamma(\frac{7}{4}-\frac{\sigma}{2})},
\end{equation}
and for a massive graviton $\Delta u(r)$ is
\begin{eqnarray}\label{deu2}
\Delta u(r)&=&{e^{-m_nr}}|\Psi_n(0)|^2\nonumber\\
&=&{e^{-m_nr}}|\frac{\Gamma(1-\sigma)}{\Gamma(-\frac{1}{4}-\frac{\sigma}{2})
\Gamma(\frac{7}{4}-\frac{\sigma}{2})}|^2.
\end{eqnarray}
Although we have obtained the parsed expression of the Newtonian potential for a massive graviton, it is cumbersome for the final result to integrate \eqref{deu2} directly. {So, our approach is to use $|\Psi_n(0)|^2$ as a fitting function with its approximate behavior,} i.e.,
\begin{equation}\label{fitting func}
|\Psi_n(0)|^2 \approx a_1~ \text{tanh}\big(a_2(m-m_0)\big)+a_3,
\end{equation}
where $a_1=\frac{26903}{100000}$, $a_2=\frac{13}{40}$, $a_3=\frac{4937}{100000}$, and $m_0=3k/2$.
The relative correction to the Newtonian potential for all the massive gravitons is
\begin{eqnarray}
\Delta U(r)&=&\int^{+\infty}_{-\infty}dz~e^{3A}
    \int_{\frac{3}{2}k}^{\infty}dm\Delta u(r)\nonumber\\
&=&b_1\frac{\pi e^{-\frac{3}{2}kr}}{kr}\Big[-b_2kr\psi ^{(0)}(b_1 k r)-b_3\nonumber\\
&&+ b_2kr\psi ^{(0)}\Big(b_1 k r+\frac{1}{2}\Big)\Big]. \label{relativeCorrectionTerm}
\end{eqnarray}
where $b_1=\frac{10}{13}$, $b_2=\frac{26903}{100000}$, $b_3=\frac{35711}{250000}$, and {$\psi^{(0)}(x)$ is the logarithmic derivative of the Gamma function: $\psi^{(0)}(x) \equiv \frac{d}{dx} \ln\Gamma(x)$}. {Note that the term in the square bracket of (\ref{relativeCorrectionTerm}) is almost a constant, we can get an approximate expression of the relative correction term of the Newtonian potential:
\begin{eqnarray}
\Delta U(r) \propto \frac{e^{-\frac{3}{2}kr}}{kr}.
\end{eqnarray}

\textbf{{Case HO:}}~The corresponding Schr\"{o}dinger equation \eqref{eq tensor} {for case HO} can be expressed as the following form
\begin{equation}
-\partial_{z}^{2}\Psi_{n}+(9k^4 z^2-3k^2)\Psi_{n}=m^2_{n}\Psi_{n}.
\end{equation}
The normalized solution is given by
\begin{equation}
\Psi_{n}=\sqrt[4]{\frac{3}{\pi}} \sqrt{\frac{1}{n!}}2^{-\frac{n}{2}} e^{-\frac{3}{2}k^2z^2} H_n\left(\sqrt{3}kz\right),
\end{equation}
where {$n$ is a positive integer} and $H_n$ is Hermite polynomial. The corresponding {mass spectrum} is $m_n=k\sqrt{6n}$, which means that the mass gap decreases with the mass. The expression of wave function $\Psi_{n}$ at $z=0$ can be rewritten as
\begin{equation}
\Psi_{n}(0)=\frac{ \sqrt[4]{3\pi }2^{n/2}}{\sqrt{n!} \,  \Gamma \left(\frac{1-n}{2}\right)}.
\end{equation}
Then, $\Delta u(r)$  for a massive graviton is
\begin{eqnarray}
\Delta u(r)&=&{e^{-m_nr}}|\Psi_n(0)|^2 \nonumber \\
 &=& {e^{-k\sqrt{6n}r}}\frac{\sqrt{3 \pi }2^n}{n! \Gamma^2 \left(\frac{1-n}{2}\right)}.
\end{eqnarray}
Note that $\Delta u(r)=0$ for an odd $n$. Therefore, the odd modes of the massive gravitons do not contribute to the correction of the Newtonian potential. In order to calculate the correction to the Newtonian potential for all the massive gravitons, we should do some tedious but simple steps. The $\Delta u(r)$  for this case can be rewritten as
 \begin{eqnarray}
\Delta u(r)&=&{e^{-m_nr}}|\Psi_n(0)|^2=
{e^{-2k\sqrt{3a}r}}\frac{\sqrt{3\pi}2^{2a}}{(2a)!\Gamma(\frac{1-2a}{2})^2}\nonumber\\
&\approx&{e^{-2k\sqrt{3a}r}}\frac{\sqrt{3}}{\pi\sqrt{a}},~~(a=n/2=\frac{m^2_n}{12k^2}),
\end{eqnarray}
However, it is cumbersome to sum $\Delta u$ directly. From Fig. \ref{jinsi}, we can see that $\sum_a\Delta u$ can be fitted by $\int\Delta u da$. For simplicity, we replace sum with integration and obtain the following approximate result
\begin{eqnarray}
\Delta U(r)&=&\sum^{\infty}_{a=1}\Delta u(r)\approx\sum^{\infty}_{a=1}{e^{-2k\sqrt{3a}r}}\frac{\sqrt{3}}{\pi\sqrt{a}}\nonumber\\
&\approx&\int^{\infty}_1 {e^{-2k\sqrt{3a}r}}\frac{\sqrt{3}}{\pi\sqrt{a}} da
=\frac{e^{-2\sqrt{3} kr}}{\pi kr}.
\end{eqnarray}
\begin{figure}[htb]
\begin{center}
\includegraphics[width=0.24\textwidth]{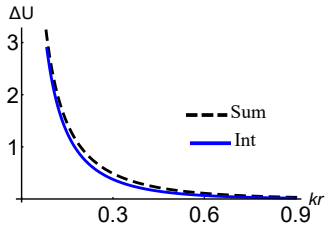}
\caption{Plot of the correction term $\Delta U(kr)$ for case HO. The dashed black line is the result of $\sum^{10000}_{a=1}\Delta u$, and the blue line the  result of $\int^{10000}_{a=1}\Delta u$.}\label{jinsi}
\end{center}
\end{figure}}

Figure \ref{correction uU} shows the correction factor $\Delta u$ of the massive gravitons (plot as $\Delta u^{1/3}$) and the relative correction terms $\Delta U$ contributed by all the massive gravitons for the three cases, respectively. From Fig. \ref{deltaum}, we can see that the correction factor $\Delta u$ is smaller for larger graviton mass, and the attenuation trends of the three cases are slightly different. {On the one hand, the different forms of $|\Psi_n(0)|^2$, the square of the massive graviton mode on the brane, are given by
\begin{equation}\label{deltaum1}
|\Psi_n(0)|^2\sim\begin{cases}
m_n/k,&\text{case~VO};\\
a_1~ \text{tanh}\big(a_2(m_n-m_0)\big)+a_3,&\text{case~PT};\\
\frac{\sqrt{3\pi}2^{n}}{n!~\Gamma\big(\frac{1-n}{2}\big)^2},~n=\frac{m^2_n}{6k^2},&\text{case~HO}.
\end{cases}
\end{equation}}
{On the other hand, for different models, the mass range of the massive gravitons which dominate the correction to the gravitational potential are different,  i.e., $(0\sim\infty)$, $[3k/2\sim\infty)$, $[2\sqrt{3}k\sim\infty)$ for the three cases, respectively.} Note that for case VO and case PT the mass spectra are continuous, while the mass spectrum of {case HO} is discrete. These differences {lead to} different forms of the relative correction terms $\Delta U$, {which are shown} in Fig. \ref{deltaUr}. From Fig. \ref{deltaUr}, we can see that on small scales $\Delta U>1$, which means that the correction term dominates the Newtonian potential, and on large scales the relative correction term  decays to zero rapidly. That is to say, the effect of all massive gravitons on the Newtonian potential can be ignored on large scales. We note that the correction term $\Delta U$ of case VO is more remarkable than the others.

\begin{figure}[htb]
\begin{center}
\subfigure[~$\Delta u(m)$]{\label{deltaum}\includegraphics[width=0.22\textwidth]{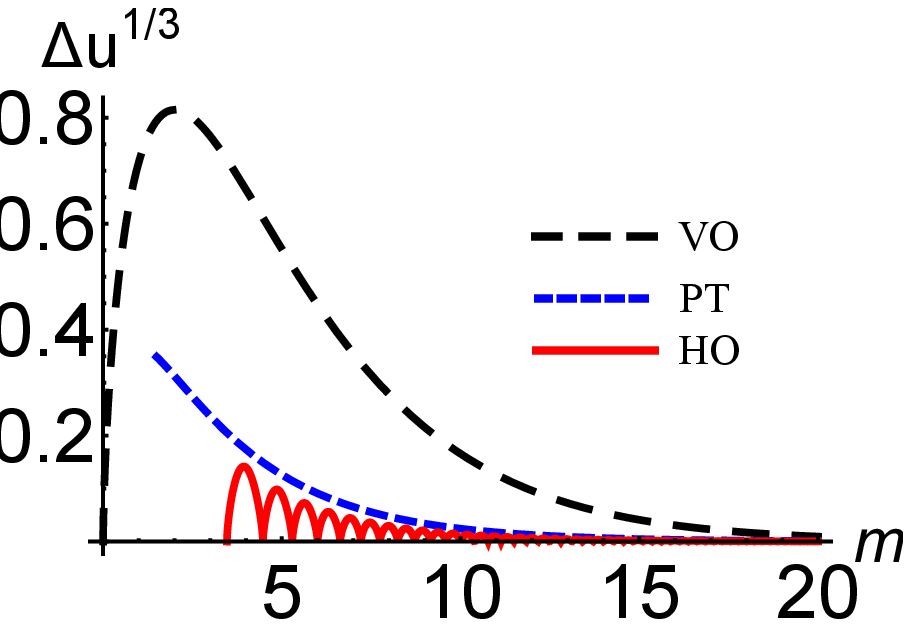}}
\subfigure[~$\Delta U(\tilde{r})$]{\label{deltaUr}\includegraphics[width=0.22\textwidth]{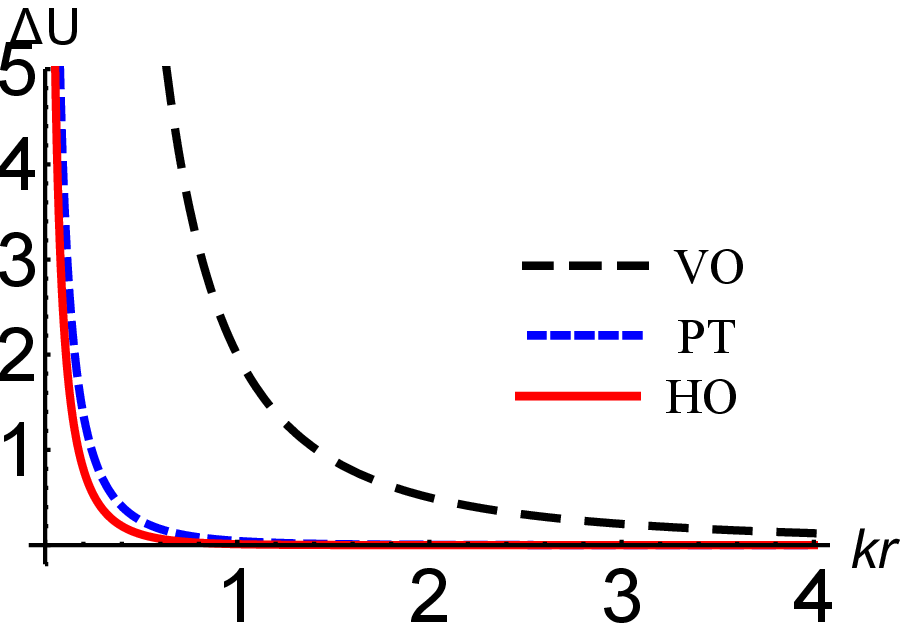}}
\end{center}
\caption{The correction factor $\Delta u$ of the massive gravitons (plot as $\Delta u^{1/3}$) and the relative correction $\Delta U$ for the three cases.  The parameter $kr$ in (a) is set as $kr=1$.}
\label{correction uU}
\end{figure}

So far, we have obtained the expressions of the three corrections to the Newtonian potential. Then, we can obtain the constraints on the parameters of our models by combining the latest tests of the gravitational inverse-square law \cite{luojun,luojun1,Lee}. In these experiments, the authors considered the following four-dimensional gravitational potential
\begin{equation}
V(r)=V_N(r)\left[1+\alpha\exp(-r/\lambda)\right],\label{potential-exe}
\end{equation}
where the parameter $r$ is the separation between two masses, $\lambda$ and $\alpha$ are the length scale and strength of the Yukawa type correction. As shown in Refs. \cite{luojun1,Lee}, the corresponding values of $\alpha, \lambda, r$ are
\begin{eqnarray}
\{\alpha,r,\lambda\}&=&\{1, 210\mu {\text{m}}, 48\mu {\text{m}}\},\label{yukawahust}\\
\{\alpha,r,\lambda\}&=&\{0.45,52\mu{\text{m}}, 38.6\mu {\text{m}}\}, \label{yukawawash}
\end{eqnarray}
 and the magnitudes of the corresponding Yukawa correction term  are
\begin{equation}\label{dUmugite}
\alpha e^{-r/\lambda}\sim\begin{cases}
0.01,~~~r=210\mu {\text{m}},\\
0.1,~~~~r=52\mu {\text{m}}.
\end{cases}
\end{equation}

Note that, the magnitude of the correction term of
the gravitational potential should be independent of its form. Therefore, it is natural to set the same magnitudes for the correction terms of our models when we choose the same separations as the separations in Refs. \cite{luojun1,Lee}. Obviously, the Yukawa type correction $\alpha\exp(-r/\lambda)$ can be considered as a form of $\Delta U$ in our models. To get the constraints on the parameters, we can set that the upper limit of the correction term $\Delta U$ is the same with the magnitude of $\alpha e^{-r/\lambda}$, which means that $\Delta U<0.01$ for $r=210\mu$m or $\Delta U\leq 0.1$ for $r=52\mu$m.

By using the above assumptions, we get the critical points  with $\Delta U(\tilde{r})=0.01$ with the separation $r=210\mu$m \cite{luojun1} and $\Delta U(\tilde{r})=0.1$ for $r=52\mu$m \cite{Lee}, where $\tilde{r}=kr$.  After calculation, we get the critical values of $\tilde{r}$ for the three cases. Note that the relative correction term $\Delta U(\tilde{r})$ decreases monotonically with $\tilde{r}$, in other words, to make sure $\Delta U$ is less than the critical points for the test experiments,  the scale parameter $k$ should satisfy the relation $k>\tilde{r}/r$. Besides, we can get the constraints on the five-dimensional fundamental scale $M_\ast$ based on Eq. \eqref{MplM2}.

We give the constraints on the parameter $k$ and the fundamental scale $M_\ast$ of our three models in Tables~\ref{initialdat1} and \ref{initialdata2}. Comparing these results, we can see that the constraints of parameters $k$ and $M_{\ast}$ based on the experimental data in Ref. \cite{Lee} are stronger than the constrains by Ref. \cite{luojun1}. {Here we should note that the specific correction to the Newtonian potential depends on the structure of brane, namely the warp factor $A(z)$ of the thick brane model. Combining Eqs. \eqref{MplM2}, \eqref{deltaum1} and Fig. \ref{correction uU}, we can see that the warp factor not only affects the mass spectrum of massive gravitons and the expression of $|\Psi_{n}(0)|^2$, but also affects the relation between $M_{pl}$ and $M_{\ast}$. In other words, the specific limits of $k$ and $M_\ast$ are also model-dependent. Therefore, comparing our three thick brane models, the limits of $k$ and $M_{\ast}$ of case VO are stricter than other two cases for both the two experimental data. Then, we can get the conclusion that the critical value of the scale parameter $k$ is at least $10^{-4}\text{eV}$, and the five-dimensional fundamental scale $M_\ast$ should be at least $10^5 \text{TeV}$.}
\begin{table*}[!htb]
\begin{center}
\caption{Constraints of the scale parameter $k$ and the {fundamental} $M_\ast$ for the three cases at $r=210\mu$m. }
\begin{tabular}{ |c | c |c |c |c |c |}
\hline
model&~$\Delta U$~&~$\tilde{r}$~&~$k_{\text{min}}(\text{eV})$~& $M_\ast$& $M_\ast^{\text{min}}(\text{TeV})$~\\
  \hline
   Case VO &$1/k^2r^2$ &14.1& $6.7\times10^{-3}$& $\frac{k}{2}M^2_{pl}$& $7.9\times10^{5}$\\
   \hline
   Case PT  & $\frac{e^{-\frac{3}{2}kr}}{kr}$&  1.6  & $7.4\times10^{-4}$&$\frac{2k}{\pi}M^2_{pl}$&$4.2\times10^{5}$\\
   \hline
   Case HO  &  $\frac{e^{-2\sqrt{3} kr}}{\pi kr}$ &  1.0  & $4.8\times10^{-4}$ &$\sqrt{3/\pi}k M^2_{pl}$ &$4.1\times10^{5}$\\
\hline
\end{tabular}
\label{initialdat1}
\end{center}
\end{table*}
\begin{table*}[!htb]
\begin{center}
\caption{Constraints of the scale parameter $k$ and the fundamental $M_\ast$ for the three cases at $r=52\mu$m.}
\begin{tabular}{ |c | c |c |c |c |c |}
\hline
\hline
model&~$\Delta U$~&~$\tilde{r}$~&~$k_{\text{min}}(\text{eV})$~& $M_\ast$& $M_\ast^{\text{min}}(\text{TeV})$~\\
  \hline
   Case VO &$1/k^2r^2$ &4.5& $8.6\times10^{-3}$& $\frac{k}{2}M^2_{pl}$& $8.6\times10^{5}$\\
   \hline
   Case PT  & $\frac{e^{-\frac{3}{2}kr}}{kr}$& 0.7 & $1.4\times10^{-3}$&$\frac{2k}{\pi}M^2_{pl}$&$5.1\times10^{5}$\\
   \hline
   Case HO  &  $\frac{e^{-2\sqrt{3} kr}}{\pi kr}$ &  0.5 & $1.0\times10^{-3}$ &$\sqrt{3/\pi}k M^2_{pl}$ &$5.3\times10^{5}$\\
\hline
\hline
\end{tabular}
\label{initialdata2}
\end{center}
\end{table*}


\section{Conclusion} \label{Sec5}
{In this paper, we considered the thick brane model generated by a mimetic scalar field with the Lagrange multiplier formulation. With the existence of excess degrees of freedom, we constructed three background solutions. Although, these three solutions have the same limit of $e^{A}|_{z\rightarrow\infty}\rightarrow 0$, the  asymptotic behaviors of these three cases are different  with $\text{HO}>\text{PT}>\text{VO}$. These different asymptotic behaviors cause different effective potentials of tensor perturbations, which lead to different corrections to the Newtonian potential by the massive KK gravitons.

We got the specific expressions of the effective potentials for the three cases. They are volcano-like potential, PT potential, and harmonic oscillator potential. We showed that all the solutions are stable under the tensor perturbations and the zero modes of tensor perturbations can be localized on the branes. Therefore, the four-dimensional Newtonian potential can be recovered.

We also calculated the corrections to the Newtonian potential for the three cases. For case VO, the relative correction term $\Delta U\propto 1/(kr)^{2}$ is the same as the RS model~\cite{Randall:1999vf}. Although, the relative correction terms $\Delta U$ of case PT and case HO have the same form with $\Delta U\approx \text{exp}(-\beta kr)/kr$, the specific values $\beta=-3/2$ and $\beta=-2\sqrt{3}$ will lead to big difference on a small scale. For the corrections to the Newtonian potential of these three cases, the results show that the four-dimensional Newtonian potential can be recovered on large scales. On a small scale, the three cases have different behaviors, the correction to the Newtonian potential of $\text{case VO}$ is more pronounced than the other two cases. {Combining the specific correction terms to the Newtonian potential for these three models and the latest experiments of the gravitational inverse-square law \cite{luojun1,Lee}, we obtained the constraints that the scale parameter $k$ is at least $10^{-4}$eV, and the corresponding five-dimensional fundamental scale $M_\ast$ should be at least $10^{5}$TeV.}

\section*{Acknowledgement}

This work was supported by the National Natural Science Foundation of China (Grants Nos. 11875151 and 11947025), the Fundamental Research Funds for the Central Universities (Grants No. lzujbky-2019-ct06), and the 111 Project under Grant No. B20063. T.T. Sui and Y.P. Zhang were supported by the scholarship granted by the Chinese Scholarship Council (CSC).

\end{document}